\documentclass[doublecol]{epl2}
\usepackage{graphicx}
\usepackage{amscd}
\usepackage{amsfonts}
\usepackage{color}
\title{Aspiration-induced reconnection in spatial public goods game
}
\shorttitle{Aspiration-induced reconnection in spatial public goods game} %Insert here a short version of the title if it exceeds 70 characters

\author{Hai-Feng Zhang\inst{1,2}\footnote{haifeng3@mail.ustc.edu.cn} \and Run-Ran Liu\inst{2}\and Zhen Wang\inst{3} \footnote{zhenwang@mail.nankai.edu.cn} \and  Han-Xin Yang\inst{2} \and Bing-Hong Wang\inst{2} \footnote{bhwang@ustc.edu.cn}}
\shortauthor{Haifeng Zhang \etal}

\institute{ \inst{1} School of Mathematical
 Science, Anhui University, Hefei 230039, P. R. China\\
  \inst{2} Department of Modern Physics, University of
Science and Technology of China, Hefei, 230026, China\\
 \inst{3} School of Physics, Nankai University, Tianjin, 300071, P. R. China
 }

 \pacs{87.23.Ge}{Dynamics of social systems} \pacs{02.50.Le}{Decision theory and game theory}
\pacs{87.23.Cc}{Population dynamics and ecological pattern formation}

\abstract {In this Letter, we introduce an aspiration-induced
reconnection mechanism into the spatial public goods game.
A player will reconnect to a randomly chosen player if its payoff acquired
from the group centered on the neighbor does not exceed the aspiration level.
We find that an intermediate aspiration level can best promote cooperation. This optimal
phenomenon can be explained by a negative feedback effect, namely, a moderate
level of reconnection induced by the intermediate aspiration level induces can change the downfall of
cooperators, and then facilitate the fast spreading of cooperation. While insufficient reconnection and excessive reconnection induced by
low and high aspiration levels respectively are not conductive to such an effect.
Moreover, we find that the intermediate aspiration level can lead to
the heterogeneous distribution of degree, which will be beneficial
to the evolution of cooperation.}

\begin{document}

\maketitle
\section{Introduction} \label{sec:intro}
Cooperation is ubiquitous in biological and social systems, yet,
understanding the emergence and maintenance of cooperative behaviors
is still a major challenge \cite{1}. In order to resolve this
puzzle, evolutionary game theory provides a fruitful framework to
address the evolution of cooperation among unrelated individuals
\cite{2,3}. For example, the prisoner's dilemma game \cite{PD1,PD2},
the snowdrift game \cite{PD1,SG} and the stag-hunt game
\cite{SH1,SH2} have been intensively studied as the paradigms to
investigate cooperative behaviors through pairwise interactions.
However, some social dilemmas involve larger groups of interactional
individuals, as can be observed by these phenomena: resource
distribution and redistribution, predator inspection behavior, alarm
calls, and group defense, health insurance, public transportation
and environmental issues. In such cases, public goods game rather
than the game of pairwise interactions seems suited to provide
reasonable explanation for the facilitation of cooperation
\cite{PGG1,PGG2,PGG3,PGGadd1,PGG4,PGG5,PGG6,PGGadd2,PGG7,Santos,PGG8}.

In the original public goods game consisting of $N$ players, each
player can decide whether to contribute an amount $c$ to the common
pool (cooperation) or not (defection). Whereafter, the overall
contributions are multiplied by a multiplication factor $r$, and
then are redistributed equally among all the players,
irrespective of their initial contributions to the common
pool \cite{PGG7}. It's obvious that selfish players are enforced
upon to select defection, which can take advantage of the public
goods. In the traditional game theory, players are perfectly
rational, thus defection becomes the dominant strategy leading to
the deterioration of cooperation, which is known as the Tragedy of
the Commons \cite{ToC1,ToC2}.

Over the past decades, a number of theoretical and experimental
evidences have been proposed and investigated to understand the
emergence of cooperation. Remarkable mechanisms include
kin selection\cite{WD}, direct and indirect
reciprocity\cite{RL,Nowak1}, punishment and reward \cite{PR1,PR2},
heterogeneous activity \cite{PGG6}, group selection \cite{AT},
voluntary participation \cite{PGG1,PGG2,PGG4}, spatial effects
\cite{Nowak2,Nowak3,PGG2}, image score effect\cite{ISE1,ISE2},
success-driven migration \cite{mobility}, to name but a few. Importantly, apart from
the considerable attention paid to the facilitation mechanisms
alone, the coevolutionary game also attracts great interest
\cite{coevolution1,
coevolution2,coevolution3,coevolution4,coevolution5,coevolution6,coevolution7,coevolution8,
coevolution9,coevolution10,coevolution11,coevolution12,coevolution13}.
Since it not only reflects the evolving of strategies over time, but
also characterizes the adaptive development of the network
topologies or the update rules (for a further review see
\cite{coevolution1}). For instance, in
\cite{coevolution2,coevolution3,coevolution4} the rewiring of
existing links was recognized as very beneficial to the evolution
of cooperation, the growth of a network had a positive impact on the
evolution of cooperation in
\cite{coevolution9,coevolution10,coevolution11}, and cooperation
could also be promoted when the coevolution of strategies and update
rules were considered \cite{coevolution12,coevolution13}. Take some
multinational corporations as examples, these corporations often
extend their business to different countries or regions to pursue
their maximal profit. However, once the profit gained from a country
or region is undesirable, they will withdraw the investment
partially or entirely, and then transfer it to other countries or
regions. Inspired by these actual phenomena and the
plentiful achievements of coevolutionary, in the Letter, we propose
an aspiration-induced reconnection mechanism to study the emergence
of cooperation in the public goods game. In the game, if player's
payoff obtained from the group centered on one of its neighbors does
not exceed the aspiration level, the player will cut the link with
the neighbor and rewire the link to one randomly selected player. 
Interestingly, we find that such a simple but meaningful approach
can promote the level of cooperation best under a moderate
aspiration level, while it is not conductive to facilitate
cooperation for too low or too high aspiration levels. We give an
interpretation to these observed phenomena by inspecting the process
of evolution and investigating the degree distribution of the
evolved network. Moreover, we examine the universality of such a
mechanism through the variation of game model.

The remainder of this paper is organized as follows. In
Sec. \uppercase\expandafter{\romannumeral2}, we will
first describe the model of considered evolutionary game. Then we
present the main results and discussions in Sec.
\uppercase\expandafter{\romannumeral3}. Finally, we will summarize
the conclusion in Sec.
\uppercase\expandafter{\romannumeral4}.

\begin{figure}
\begin{center}
\includegraphics[height=6cm,width=80mm]{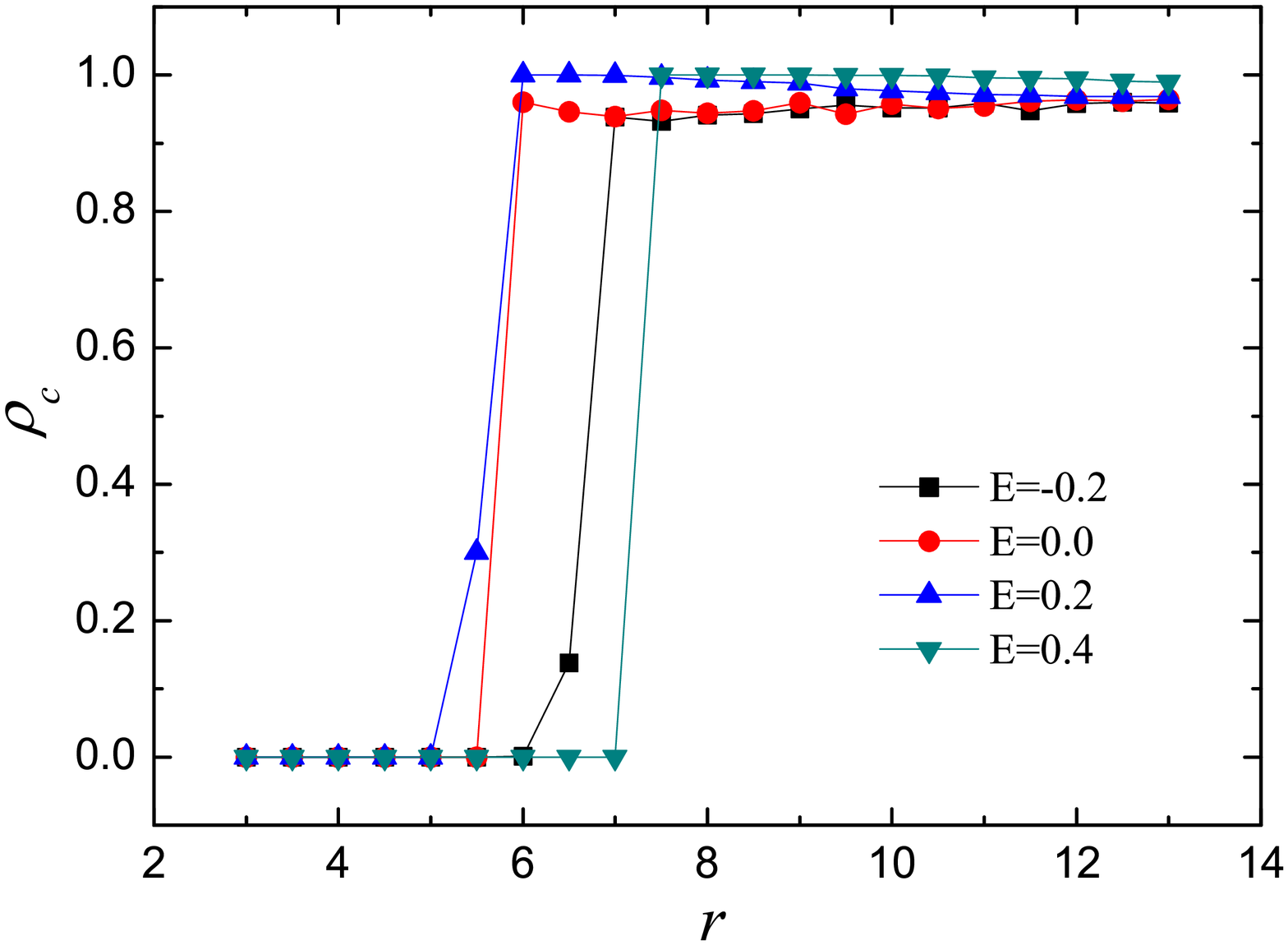}
\caption{((Color online) Fraction of cooperators $\rho_c$ in
dependence on the multiplication factor $r$ for different values of
aspiration level $E$. Note that intermediate value of $E$
can sustain cooperation better than lower or larger case.}
\label{fig1}
\end{center}%
\end{figure}

\section{Evolutionary game} \label{sec:model}

We consider the public goods game with players located on the nodes
of the spatial network. According to Ref. \cite{Santos},
each individual $i$ participates in interactions in $k_i+1$
neighborhoods that center about $i$ and its $k_i$ neighbors, where
each neighborhood contains a central node and all nodes that are
directly connected to it. Each cooperator contributes a total cost
$c=1$ shared equally among all the neighborhoods that it engages.
The strategy is $s_x=1$ for a $C$ player and $s_x=0$ otherwise. The
payoff of the individual $x$ with strategy $s_x$ associated with the
neighborhood centered at an individual $y$ is given by
\begin{equation}\label{eq1}
   p_{x,y}=\frac{r}{k_y+1}\sum_{i=0}^{k_y}\frac{s_i}{k_i+1}-\frac{s_x}{k_x+1},
\end{equation}
where $i=0$ represents $y$, $s_i$ is the strategy of the neighbors
$i$ of $y$, and $k_i$ is its degree. The total payoff of player $x$
is
\begin{equation}\label{eq2}
  P_x=\sum_{y\in\Lambda_x}p_{x,y},
\end{equation}
where $\Lambda_x$ is the neighborhood of $x$ and itself
\cite{PGG7}.

\begin{figure}
\begin{center}
\includegraphics[height=8cm,width=80mm]{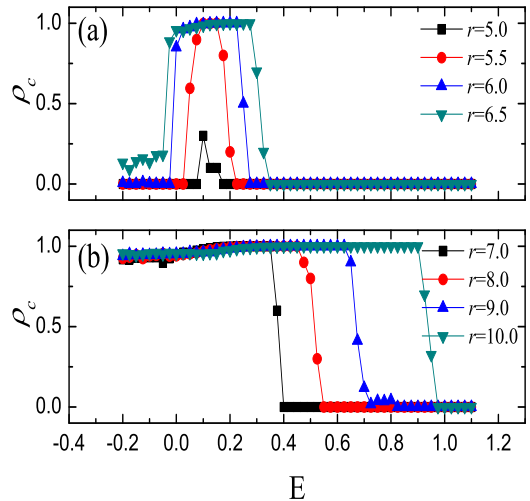}
\caption{(Color online) Fraction of cooperators $\rho_c$ in
dependence on aspiration level $E$ for different values of
multiplication factors $r$. Note that for $r\leq6.5$ (a), it can be
obviously observed that there exists an intermediate aspiration
level $E$ for which the evolution of cooperation is optimal; while
for $r\geq7$ (b), this peak is not distinct as the above case
because of the fact that large $r$ can promote cooperation better,
even though the invariance of network.} \label{fig2}
\end{center}%
\end{figure}

As the interaction network, we use a Newman-Watts small world
network, where a number of long-range links $N_{add}$ are randomly
added to the two-dimensional lattice with periodic boundary
conditions \cite{NW}. Initially, each player is designed
to be either a cooperator or defector with equal probability.
Players asynchronously update their strategies in a random
sequential order. Before updating strategy, the randomly selected
player $x$ evaluates the payoff $p_{x,y}$ from the neighborhood
centered on $y$ according to Eq. (\ref{eq1}). If the payoff
$p_{x,y}$ cannot satisfy the aspiration level $E$ ($E$ is
uniform for all the players), player $x$ will remove the link with
neighbor $y$, and then create a new link to a randomly chosen non-neighbor node from network
(i.e., multiple links are prohibited.) In our model, we assume that the
local interactions with four nearest neighbor nodes (the von Neumann
neighborhood) do not engage in the reconnection behaviors. It is
because that these local interactions are determined by the spatial
neighborhoods, and they are reasonably assumed to be fixed during
the whole process \cite{coevolution5}. After rewiring, player $x$
collects its total payoff according to Eq. (\ref{eq2}), then all the
neighbors of player $x$ acquire their payoffs by means of the same
way as player $x$.

Lastly, the player $x$ will randomly select one of its neighbors $z$
and adopts $z$'s strategy with a probability $W(s_{z} \to s_{x})$
depending on the payoff difference. Namely

\begin{eqnarray}
 W(s_{z}\to s_{x}) &=&\frac{1}{1+exp[(P_{x}-P_{z})/K]
 },
\end{eqnarray}
where $K$ denotes the amplitude of noise or its inverse ($1/K$), the
so-called intensity of selection \cite{noise}. In the limit $K \to
0$, the strategy of neighbor $y$ is always adopted provided that
$P_{y}>P_{x}$. While in the limit $K \to \infty$ all information is
lost, that is, player $x$ switches to the strategy of neighbor $y$
by tossing a coin. Furthermore, it should be noted that
this update rule is the sequential version of the Fermi rule, at
variance with some other works in which updates are made in
parallel \cite{PGG7,PGG8}. Since the effect of $K$ on the evolution of
cooperation has been studied in detail in
Refs.\cite{PGG8,noise1,noise2}, we simply set the value of $K$ to be
$K=0.1$.

%%Following previous studies \cite{noise1,noise2}, we set $\kappa=0.1$.

\section{Results and discussions} \label{sec:main results}

Results of Monte Carlo simulations presented below are obtained on a
Newman-Watts small world network hosting $N=2500$ players with
average degree $\langle k\rangle=8$, namely, the number of
long-range links is $N_{add}=2N$ (By extensive simulations, we find
our results are robust to the network sizes and the
average degrees.). On average, in each Monte Carlo step
(MCS), all players have the chances of rewiring and updating their
strategies. The key quantity for characterizing the cooperative
behavior is the fraction of cooperators $\rho_c$, which is gained by averaging 2000
full steps after a transient time of 10000 generations in our work. Moreover,
since the process of rewiring may yield heterogeneous distribution
of links, which will seriously affects the accuracy of simulations.
Therefore, the final results are averaged over 40 independent runs
for each parameter value.

\begin{figure}
\begin{center}
\includegraphics[height=77mm,width=77mm]{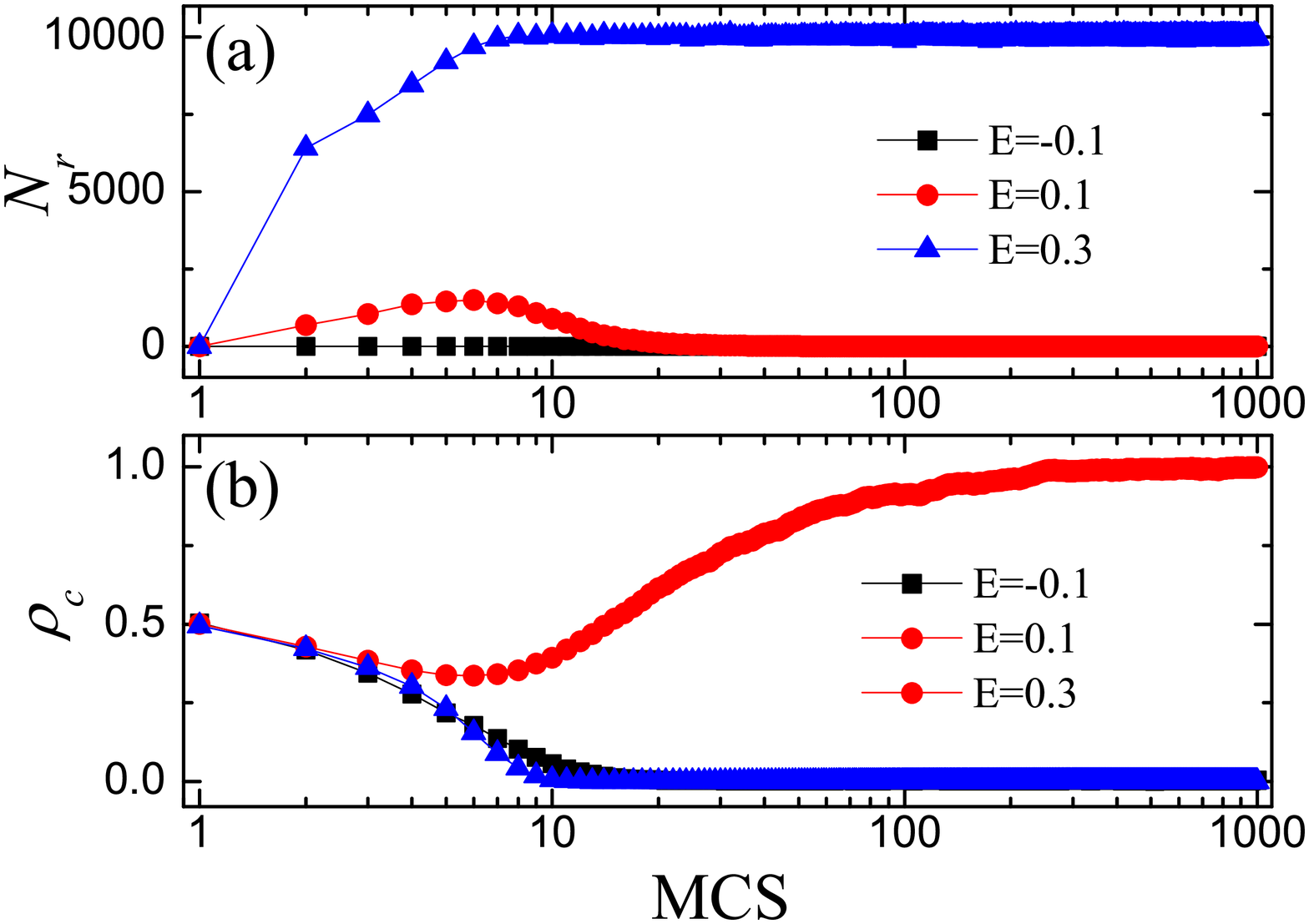}
\caption{(Color online) Time courses depicting the number of
reconnections $N_r$ and the fraction of cooperators $\rho_c$ for
different values of $E$. Note in (a), the number of reconnections will
increase at first for intermediate and large aspiration levels, but
develops into different patterns finally. While for small aspiration
level, the number of reconnections will keep constant. For the evolution of
cooperation in (b), the intermediate aspiration level will go through a
negative feedback effect to promoting cooperation, which corresponds
to a peak of reconnections, while small and large aspiration levels
can not avoid the destiny of cooperation dying out. All the results
are obtained for $r=6.0$. }\label{fig3}
\end{center}%
\end{figure}

Figure \ref{fig1}
shows the fraction of cooperators $\rho_c$ as a function of the
multiplication factor
 $r$ for different aspiration levels $E$. One can see that, for
 $E=0.0$ or $E=0.2$, the cooperative phenomenon can be flourished even when
 $r\leq 5.5$ (especially, for the case of $E=0.2$, $\rho=1$ when $r=5.5$), yet for $E=-0.2$ or $E=0.4$, cooperative phenomenon appears only for $r\geq 6$.
 These results suggest
that there may exist an intermediate aspiration level $E$, which induces the optimal cooperation.
To examine the effect of $E$ on the evolution of cooperation more
precisely, we present $\rho_c$ in dependence on the aspiration level
$E$ for different values of $r$ in Fig.~\ref{fig2}. As
shown in Fig. \ref{fig2}(a), the impact of
small $E$ on cooperation remains marginal, and thus most cooperators
could not resist the exploitation against defectors or only a small
fraction of cooperators could survive.  However, as
aspiration level $E$ reaches an intermediate value, a remarkable
increase of cooperation can be observed. By further increasing the
aspiration level, the facilitation effect is deteriorated
again. These results favor that the intermediate aspiration level
can warrant an optimal promotion of cooperation, which is analogous
to the so-called coherence resonance \cite{resonance1,resonance2}.
For large $r$ (i.e., $r\geq7$), as shown in Fig.
\ref{fig2}(b), the non-monotonic phenomenon is not so distinct as
the case of small $r$. This is actually what one would expect,
because, compared with small $r$ where the sum of cooperators is
very limited or zero, large $r$ can make sure of generating more
rewards for the cooperation, even though the network does not
evolve.

\begin{figure}
\begin{center}
\includegraphics[height=6cm,width=80mm]{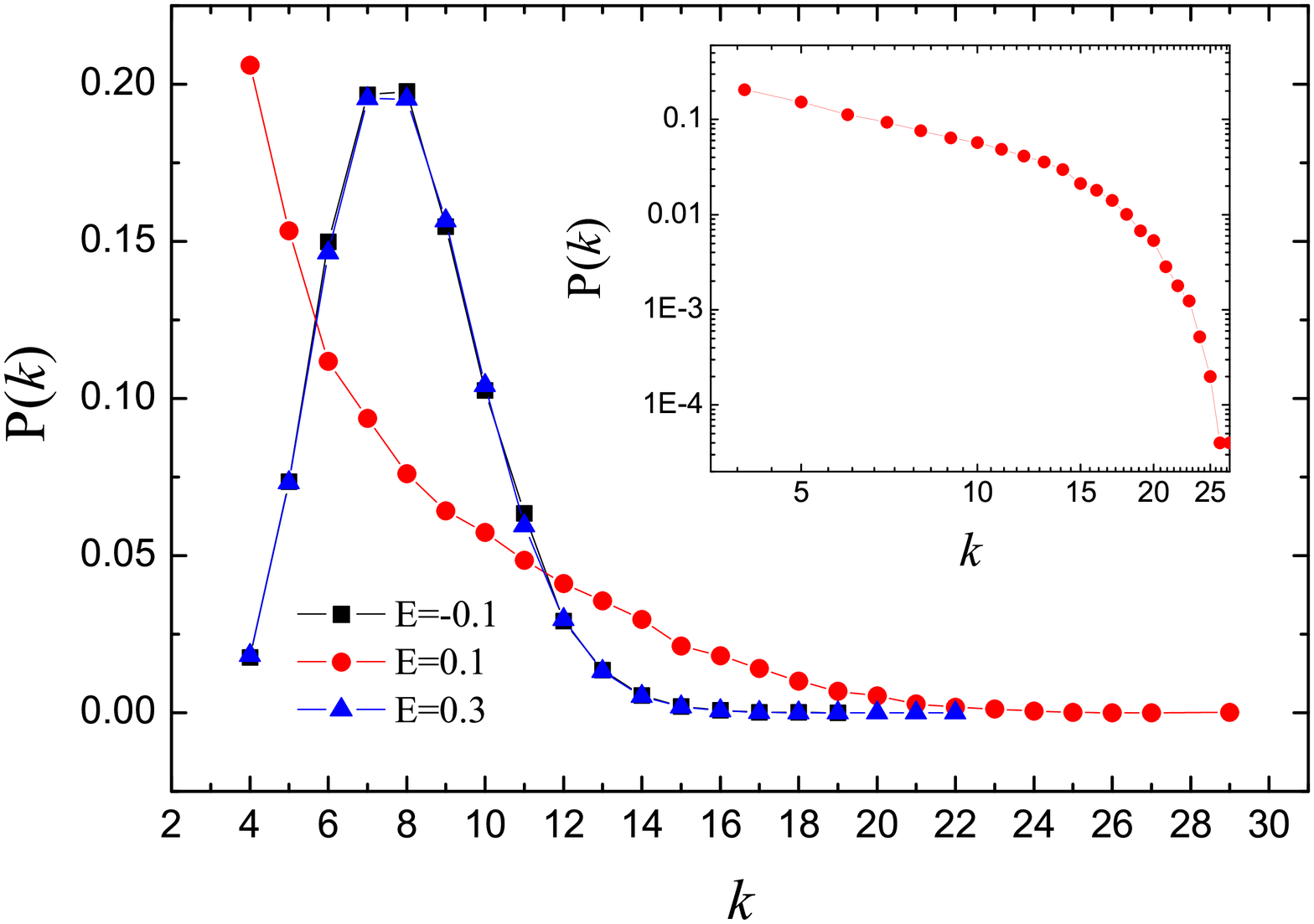}
\caption{(Color online) Degree distributions $P(k)$ of networks for
different aspiration levels $E$ in the final steady states. Note
that small and large aspiration levels will not alter the initial
norm distribution with $\langle k\rangle=8$, while intermediate
aspiration level induces an explicit heterogeneous distribution that
is more beneficial for the evolution of cooperation 
(see also the inset for $E=0.1$). All the results
are obtained as averages over 50 independent realizations for
$r=6.0$. }\label{fig4}
\end{center}%
\end{figure}

In order to understand the nontrivial effect of aspiration
level on the fraction of cooperators, we study the time courses of
reconnection number $N_r$ and the fraction of cooperators $\rho_c$ for
different values of $E$ in Fig.~\ref{fig3}. For $E=-0.1$, Figure~\ref{fig3}(a) illustrates that due to the fact that individual
payoffs are always higher than aspiration level, the initial
topology structure of interaction network does not change over time,
namely, the number of reconnections always equals zero. At this time,
cooperators on the initial network cannot resist the exploitation
from defectors, which results in the extinction of cooperation (see
Fig.~\ref{fig3}(b)). Whereas for large aspiration level ($E=0.3$),
most individuals' payoffs are lower than the aspiration level. As a
result, reconnection behavior occurs frequently and individuals will randomly
connect to others as time evolves. In such case, the cooperative
clusters are fragile and are easy to be destroyed by the invasion of
the defectors. So large aspiration level also induces the extinction
of cooperation. More significantly, we observe that the immediate
aspiration level ($E=0.1$) can induce a peak of reconnections which
yet corresponds to a downside for the evolution of cooperation. In
the most early stages of process, defection yields higher individual
benefit, and the outlook for cooperators is gloomy. With the game
forward, a few players' payoffs could not exceed the aspiration
level, and thus a small fraction of reconnections will appear. All
the reconnections will effectively alter the evolution tide, namely,
the downfall of cooperators will transform into the fast spreading
of cooperators. Because, in such case, only choosing cooperation
adequately can exceed the aspiration level. When cooperation becomes
the dominant strategy, individual payoffs will uniformly exceed the
aspiration level, which makes the possibility of reconnection
vanish, namely, the number of reconnections will become zero over
again. Consequently, an intermediate aspiration level can result in
a peak of reconnections which promotes the level of cooperation
by means of a negative feedback effect. While for low or high
aspiration levels, insufficient or excessive reconnections could not
provide advantageous conditions for such a feedback effect.

\begin{figure}
\begin{center}
\includegraphics[height=8cm,width=80mm]{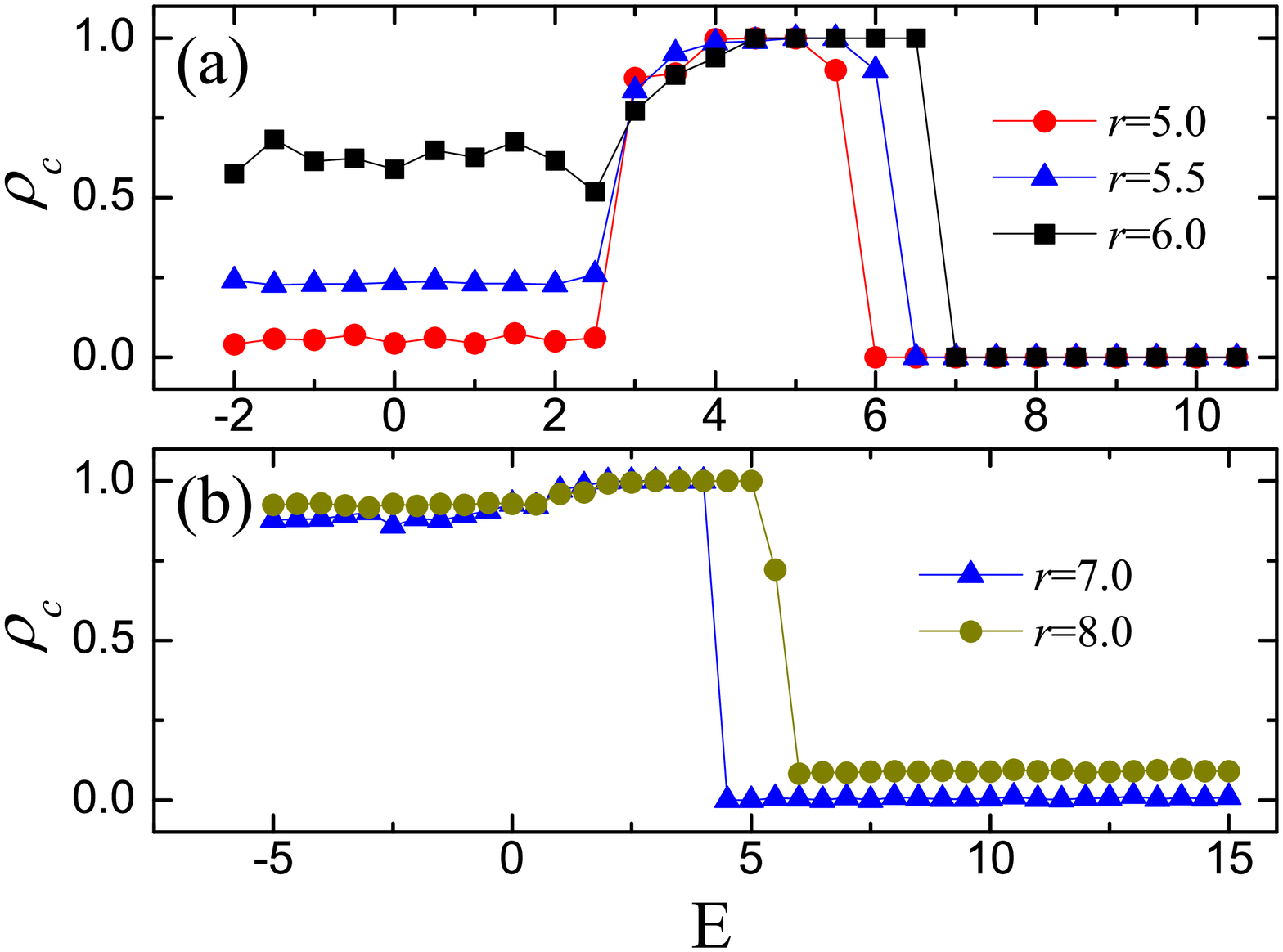}
\caption{(Color online) Fraction of cooperators $\rho_c$ in
dependence on aspiration level $E$ for different values of
multiplication factor $r$ in the game where every cooperator
separately contributes a cost $c=1$ to each neighborhood that it
engages. Note that the optimal promotion phenomena of cooperation
are similar to those presented in Fig. \ref{fig2}.}\label{fig5}
\end{center}%
\end{figure}

Subsequently, it remains of interest to examine the degree
distributions of the evolved networks for different aspiration
levels. It has been known that the heterogeneity of host network
plays an important role in the substantial promotion of cooperation.
If the hub node takes the cooperation (defection) strategy, its
strategy will become an example to be imitated by its neighborhood,
which increases (decreases) cooperators' (defectors') payoffs and
results in the great promotion of cooperation
\cite{PD1,PGG7,Santos}. Hence, what we would expect is to present a
highly heterogeneous distribution as well. Figure~\ref{fig4}
demonstrates clearly the degree distributions in the final steady
states. According to the time courses of $N_r$ in Fig.~\ref{fig3}(a), we observe that the network does not evolve with time
for $E=-0.1$, so the degree distribution obeys norm distribution
with $\langle k\rangle=8$ (see Fig.~\ref{fig4}). While for $E=0.3$,
long-range links which are generated randomly and excessively make
the degree distribution confirm to the norm distribution likewise.
 Interestingly, for $E=0.1$, the degree distribution
deviates from the norm distribution and the heterogeneity of network is
formed (see also the inset of Fig.~\ref{fig4} by using a log-log representation), which
is consistent with our expectation. The highly heterogeneous
distribution induced by the intermediate aspiration level is crucial
for the optimal promotion phenomenon of cooperation presented in
Fig.~\ref{fig2}. Therefore, the promotion of cooperation partly
attributes to the potential heterogeneous states within the network.

In the above evolutionary game, we assume that each cooperator
contributes the same total cost $c=1$, which is then shared equally
among all the neighborhoods that it engages in, namely, the individual
contribution is independent of the number of its social ties.
Whereas, in the opposite limit considered in Ref. \cite{Santos},
every cooperator separately contributes a cost $c=1$ to each
neighborhood that it engages. In this case, the total contribution
of each cooperator $x$ is equal to $k_x+1$. Similarly as in Eq.
(\ref{eq1}), the payoff of player $x$ (with strategy $s_x$) obtained
from the neighborhood centered on the player $y$ is given by

\begin{equation}\label{eq3}
   p_{x,y}=\frac{r}{k_y+1}\sum_{i=0}^{k_y}s_i-s_x.
\end{equation}

Interestingly, as shown in Fig. \ref{fig5}, there also exists an
intermediate aspiration level, leading to the highest cooperation
level when $r$ is fixed. Thus, the aspiration-induced reconnection
mechanism is robust for promoting cooperation, regardless of the
total contribution of cooperators.

\section{Summary} \label{sec:discussion}

In summary, we have studied the effect of the aspiration-induced
reconnection on cooperation in spatial public goods game. In the
game, if player's payoff acquired from the group centered on the
neighbor does not exceed aspiration level, it will remove the link
with the neighbor and reconnect it to a randomly selected player.
Through scientific simulations, we have shown that, irrespective of
the total contribution of each cooperator, there exists an
intermediate aspiration level results in the optimal cooperation. This optimal phenomenon can be explained by means of
a negative feedback effect. In the early stages of evolutionary
process, though cooperators are decimated by defectors, the peak of
reconnection induced by the intermediate aspiration level will
change the downfall of cooperation and facilitate the fast spreading
of cooperation. While for too low or too high aspiration levels,
insufficient or excessive reconnections does not provide any
possibility for the emergence of such a feedback effect. Moreover,
we have analyzed the degree distributions of the network. 
Of particular interest is that the heterogeneous degree distribution
induced by intermediate aspiration level, warrants a potent
promotion of cooperation. Since the phenomena related with
aspiration-induced reconnection are abundant in our society, we hope
that our results can offer a new insight into understanding these
phenomena.

\begin{acknowledgments}
This work is funded by the National Natural Science Foundation of
China (Grant No. 10975126, 10635040,11047012) and the Specialized Research
Fund for the Doctoral Program of Higher Education of China (Grant
No. 20060358065). HZ is funded by the 211 Project of Anhui
University(2009QN003A, KJTD002B), and the National Natural Science
Foundation of China (Grant No. 11005001). ZW is funded by the Center
for Asia Studies of Nankai University (Grant No. 2010-5) and the
National Natural Science Foundation of China (Grant No. 10672081 and
Grant No. 60904063).
\end{acknowledgments}

\end{document}